\documentclass{article}
\usepackage[top=1.5in, bottom=1.5in, left=1.2in, right=1.2in]{geometry}
\usepackage{authblk}
\usepackage{amssymb, amsmath,framed}
\usepackage{graphicx}
\usepackage[round]{natbib}

\usepackage{bm}
\usepackage[colorlinks=true,linkcolor=blue,urlcolor=blue,citecolor=blue,anchorcolor=blue]{hyperref}
\expandafter\ifx\csname package@font\endcsname\relax\else
 \expandafter\expandafter
 \expandafter\usepackage
 \expandafter\expandafter
 \expandafter{\csname package@font\endcsname}
\fi
\def\bq{\begin{equation}}
\def\eq{\end{equation}}
\def\bqy{\begin{eqnarray}}
\def\eqy{\end{eqnarray}}

\makeatother

\begin{document}

\title{Planetary Scale Information Transmission in the Biosphere and Technosphere: Limits and Evolution}

\author{Manasvi Lingam\thanks{Electronic address: \texttt{mlingam@fit.edu}}}
\affil{Department of Aerospace, Physics and Space Sciences, Florida Institute of Technology, Melbourne, FL 32901, USA}
\affil{Department of Physics and Institute for Fusion Studies, The University of Texas at Austin, Austin, TX 78712, USA}

\author{Adam Frank}
\affil{Department of Physics and Astronomy, University of Rochester, Rochester, NY 14620, USA}

\author{Amedeo Balbi}
\affil{Dipartimento di Fisica, Universit\`a di Roma ``Tor Vergata", 00133 Roma, Italy}

\date{}

\maketitle

\begin{abstract}
Information transmission via communication between agents is ubiquitous on Earth, and is a vital facet of living systems. In this paper, we aim to quantify this rate of information transmission associated with Earth's biosphere and technosphere (i.e., a measure of global information flow) by means of a heuristic order-of-magnitude model. By adopting ostensibly conservative values for the salient parameters, we estimate that the global information transmission rate for the biosphere might be $\sim 10^{24}$ bits/s, and that it may perhaps exceed the corresponding rate for the current technosphere by $\sim 9$ orders of magnitude. However, under the equivocal assumption of sustained exponential growth, we find that information transmission in the technosphere can potentially surpass that of the biosphere $\sim 90$ years in the future, reflecting its increasing dominance.
\end{abstract}

\section{Introduction}\label{SecIntro}

It is widely accepted that information processing constitutes a fundamental aspect of life \citep[e.g.,][]{Sch44,MC96,SS00,GF07,PN08,DCK11,CA12,HZ13,DRT13,FNG13,DI16,LT16,TB16,SU20,CS21,FL21,LiM21,LKAL,TG21,FGW22}. In this ``informational'' realm, the significance and ubiquity of information transmission between organisms/agents via appropriate communication channels is thoroughly documented, ranging from microscopic to macroscopic scales and spanning multiple modalities  \citep[e.g.,][]{RCR50,WRL99,DF09,DBL10,YB12,BL16,WBM18,MTB19,MS21,JAS21,EKB23}. 

A number of authors have hypothesized that exploring information transmission via communication might be valuable for understanding various major evolutionary events in Earth's history. The proposals have ranged from abiogenesis and early evolution of (proto)life \citep{MS97,WRL99,MTB19,VW19,GW20,EKB23} to breakthroughs in cooperation, collective behavior, multicellularity and cognition \citep{JAS98,BCL00,BJC01,EBJ09,PL15,ML18,JAS21,LKAL}. Even setting aside these concepts, comparisons of key properties of Earth's biosphere and ``technosphere'' (see, e.g., \citep{VIV98,ZWW17}) can yield potentially useful insights into how much the latter is catching up with, or impinging, on the former, which is increasingly timely in the Anthropocene \citep{LM18,TWZ20}. Comparisons of aspects of the biosphere and the technosphere have recently been published in the context of mass \citep{EBG20} and internal information processing \citep{GHK18}.

Hence, it is natural to wonder about the aforementioned rate of information transmission (measured in bits/s) ascribable to modern Earth's biosphere and its technosphere, at a given instant in time, which follows from generalizing this notion to encompass the planet \emph{in toto} -- the overarching theme of this paper. Analyzing the global rate of information transfer may shed light on the communication load embedded in planetary-scale biological and technological systems. This theme is important both from the standpoint of understanding the genesis and evolution of these systems on Earth, as well as assessing the astrobiological possibility of their emergence elsewhere \citep{SMI18,LL19,Co20,ML21}.

Due to the manifold uncertainties at play (as illustrated hereafter), we report a simple order-of-magnitude estimate (Fermi problem) in the spirit of the famous Drake equation from astrobiology \citep{Drake65,ML19,WHF22}. We will first examine the global rate of information transmission in the biosphere (Section \ref{SSecITBio}), and then delve into the equivalent calculation for the human-mediated technosphere (Section \ref{SecITTechno}). Finally, we articulate the ramifications of our preliminary findings in Section \ref{SecConc}.

\section{Information transmission in the biosphere}\label{SSecITBio}

We shall focus on prokaryotes in this analysis (denoted by the subscript `$p$'), whose communication is modulated by signaling molecules \citep{NEH,PKTG}. The total biomass of eukaryotes is probably $\sim 6$ times greater than that of prokaryotes \citep[Figure 1]{BPM18}, but the mean volume, and consequently the average mass, of a single eukaryotic cell is $\sim 10^4$ times greater than its prokaryotic counterpart (refer to \citet{CD96} and \citet{MP16}). Hence, the total number of prokaryotic cells on Earth may be approximately three orders of magnitude higher than the equivalent for eukaryotic cells; the relevance of this inference will shortly become apparent.

The total number of prokaryotes is $N_p$, and the mean number of prokaryotes with which a single prokaryote can communicate is $k_p$; we shall conservatively assume that prokaryotes convey information to merely their immediate neighbors. Therefore, the total number of communicating pairs that could exist in principle is $\sim N_p k_p$. However, of the total population of prokaryotes, only a fraction $f_p$ (which may perhaps approach unity in some settings) are active and taking part in molecular communication at an arbitrary moment in time. The global rate of information transmission attributable to the prokaryotic realm (denoted by $I_p$) is accordingly estimated to be
\begin{equation}\label{Ipdef}
    I_p \sim N_p \cdot k_p \cdot f_p \cdot \mathcal{S}_p,
\end{equation}
where $\mathcal{S}_p$ represents the information transmission rate between a characteristic pair of prokaryotes, conventionally expressed in terms of the mutual information \citep{LN14}. It is reasonable to concentrate on biofilms henceforth, since a near-majority of all prokaryotes (circa $40$-$80\%$) on our planet are presumed to be residents of biofilms \citep{FW19}. 

We specify $N_p \sim 7 \times 10^{29}$ \citep{WCW98,FW19},\footnote{This value is constructed by taking the estimate of $1.2 \times 10^{30}$ cells from \citet{FW19}, and multiplying it with a biofilm fraction of $60\%$.} along with $k_p \sim 6$ \citep[Figure 1D]{LZK18},\footnote{This choice is identical to the number of nearest neighbors manifested in the efficient method of hexagonal close packing in two dimensions.} for prokaryotes occurring in biofilms, although the latter could be enhanced under optimal conditions \citep{DPL17,MGB20}. Reliable constraints on $f_p$ do not seem to be available, especially for the substantial deep biosphere, which remains scarcely investigated. We will suppose that $\sim 40\%$ of all cells are active at any moment \citep{JL10,RBS13},\footnote{In actuality, even dormant microbes (in the form of spores) are capable of information sensing, and responding to environmental signals \citep{KGW22}, but we neglect their relative contribution here.} and that the above subset of active cells participate in signaling and communication merely around $1\%$ of the time; the latter may constitute a strongly conservative assumption in some settings (consult, e.g., \citet{LZK18}). The preceding fiducial values collectively translate to selecting $f_p \sim 4 \times 10^{-3}$. 

Finally, we turn our attention to $\mathcal{S}_p$ for prokaryotes. \citet{MKME} tackled \emph{Escherichia coli} chemotaxis via experiments and theoretical modeling, and determined that information could be acquired at rates up to $\sim 10^{-2}$ bits/s, but specifically in environments with shallow chemical gradients (length scales of mm to cm). A theoretical study of \emph{E. coli} chemotaxis by \citet{TT09} arrived at a similar rate, but entailed simplifying assumptions like Gaussian white noise. In a clearly different context, wherein the goal was to effectuate an experimental molecular communication system with an optical-to-chemical signal converter using \emph{E. coli}, \citet{GKP18} reported an information transmission rate close to $10^{-2}$ bits/s.

On the other hand, \emph{E. coli} engineered to transmit data (i.e., functioning as molecular communication systems) via modulation schemes such as ``time-elapse communication'' have been shown to yield information transmission rates of $\sim 10^{-4}$ bits/s \citep{KAB}. Although we focus strictly on intercellular signaling, we remark in passing that \emph{intracellular} signaling via the ethylene signal transduction pathway in the well-known plant \emph{Arabidopsis thaliana} might also exhibit information transmission rates of this order. On the basis of this discussion, we adopt a communication rate of $\mathcal{S}_p \sim 10^{-4}$ bits/s; however, we caution that $\mathcal{S}_p$ could be even smaller for certain prokaryotes.

On substituting the prior values into (\ref{Ipdef}), we obtain a global information transmission rate of $I_p \sim 1.7 \times 10^{24}$ bits/s for the modern biosphere. This rate, at first glimpse, is many orders of magnitude smaller than $\sim 10^{39}$ bits/s, which is an optimistic estimate for the proposed computational speed of the biosphere \citep{LFC15}. An approximate analog for the technosphere may be global computing power, which is anticipated to potentially reach $\sim 10^{23}$ FLOPS (floating point operations per second) in 2030 if artificial intelligence (AI) computing is taken into account, as per the \emph{Intelligent World 2030} report from Huawei \citep[e.g.,][]{MWS23},\footnote{\url{https://www.huawei.com/uk/giv}} and other informal studies.\footnote{\url{https://www.diskmfr.com/this-article-is-worth-reading-about-the-computing-power/}}

The discrepancy between $I_p$ and the above computational speed of the biosphere is explained by the fact that \citet{LFC15} were primarily focused toward quantifying the transcription rate of nucleotides as a proxy for computation, whereas our emphasis is manifestly on information shared between agents via communication channels. An emphasis on information {\it transfer} foregrounds the networked nature of the biosphere's dynamics, which is arguably as important as analyzing the typically self-contained nature of individual DNA replication events. 

At this juncture, some key caveats pertaining to (\ref{Ipdef}) are necessary, which also apply to the subsequent (\ref{Ihdef}), albeit to a lesser degree. First, we reiterate that (\ref{Ipdef}) comprises an order-of-magnitude approach that does not reflect, among other effects, spatial heterogeneity and temporal evolution. Second, we see that (\ref{Ipdef}) has a linear dependence on four factors, and each has some uncertainty associated with them. The degree of variation attributable to $N_p$ and $k_p$ is not substantial because biofilms have been widely scrutinized, but $f_p$ and $\mathcal{S}_p$ are not tightly constrained. However, as the scaling is linear (and not a higher power-law exponent), our results may not be greatly affected if the estimates provided for these variables are not incorrect by orders of magnitude.

Before moving onward, a brief comment on eukaryotes is warranted. We would need to replace all the parameters in (\ref{Ipdef}) with their eukaryotic equivalents (labeled by the subscript `$e$'), in the case of intercellular communication in eukaryotes. However, because $N_p \sim 10^3 N_e$ is expected, as outlined earlier in Section \ref{SSecITBio}, there are far more prokaryotes than eukaryotes. Hence, if all other factors are held equal in (\ref{Ipdef}), the information transmission between prokaryotes would dominate that of eukaryotes. In other words, $I_e > I_p$ may only be attainable when the remaining variables in (\ref{Ipdef}) are collectively several orders of magnitude higher for eukaryotic cells relative to prokaryotes; however, exact or even approximate values for eukaryotes are not easy to discern. Even if this scenario (of $I_e > I_p$) were to be applicable, it is still justified to interpret $I_p$ as a plausible lower bound for the information transmission rate associated with the biosphere, because the contribution from eukaryotes is not included by definition in $I_p$.

\section{Information transmission in the technosphere}\label{SecITTechno}

In terms of the evolution of informational dynamics of the biosphere, eukaryotes have played a progressively pivotal role. In particular, one eukaryotic species has shaped Earth profoundly, to wit, \emph{Homo sapiens}. This pivot to highlighting humans is reminiscent of the comparison of biomass and human-made mass; the latter potentially exceeded the former in the year 2020 \citep{EBG20}. The technosphere engendered by anthropogenic activity has witnessed tremendous growth \citep{ZWW17}, especially so in the Anthropocene \citep{LM18,TWZ20}. If we consider humans as the ``units'' in lieu of prokaryotes and introduce the subscript `$h$', the analogue of (\ref{Ipdef}) for human-to-human communication is duly expressible as
\begin{equation}\label{Ihdef}
    I_h \sim N_h \cdot k_h \cdot f_h \cdot \mathcal{S}_h.
\end{equation}

We input $N_h \approx 8 \times 10^9$ for the current human population. Next, the information transmission rate for $17$ human languages drawn from $9$ language families was estimated by \citet{COD19} to equal $\mathcal{S}_h \approx 39$ bits/s. The majority of human-to-human communication likely involves small groups, although a fraction of interactions (e.g., public events) are much larger; we adopt an average value of $k_h \sim 10$ drawing on these considerations. Finally, as most humans are anticipated to actively communicate for a few hours per day, we will select $f_h \sim 0.1$. After plugging these choices in (\ref{Ihdef}), we obtain a global information transmission rate of $I_h \sim 3.2 \times 10^{11}$ bits/s. On comparison with $I_p$, we notice that $I_h$ is over $12$ orders of magnitude lower.

However, an inherent subtlety vis-\`a-vis the above discussion merits explication. The technosphere in the Anthropocene evinces escalating information flow \citep{ZWW17}, which extends far beyond direct human-to-human communication (tackled in the prior paragraph) and encompasses the Internet, among other communication channels \citep{HL11}. The advent of the Internet of Things (IoT), entailing Machine-To-Machine (M2M) connections, is swiftly accelerating this trend \citep{AIM10,HTM14,SG21}, with M2M connections probably comprising half of the total in 2022-23 \citep{Cisco}.\footnote{\url{https://www.cisco.com/c/en/us/solutions/collateral/executive-perspectives/annual-internet-report/white-paper-c11-741490.html}} Hence, it is worth assessing the global rate of information transmission via human-made networks (denoted by $I_t$); in this context, Internet traffic can be singled out because of its pace of development.

The Internet traffic consists of various components, some of which (e.g., mobile networks) are expanding at faster rates than others. We employ consumer Internet Protocol (IP) traffic as the proxy. In \citet{Cisco}, the worldwide IP traffic was estimated to be $1.5 \times 10^{21}$ bytes per year, which translates to $3.8 \times 10^{14}\,\mathrm{bits/s}$. The projected growth rate in \citet{Cisco} was $26\%$ per year during the time frame of 2017 to 2022, which is equivalent to an $e$-folding timescale of $4.33$ years. Interestingly, the above growth rate displays excellent agreement with the telecommunication growth rate of $28\%$ in the period of 1986 to 2007 determined in a comprehensive analysis by \citet{HL11}, after tracking $60$ analog and digital technologies across this time span.

While the above paragraph indicates that a growth rate of approximately $26\%$ might be a reasonable assumption, we emphasize that this rate can experience either an increase or decrease over the span of decades. For example, technological breakthroughs could facilitate a boost, certain natural or artificial disasters that destroy infrastructure may lower the growth rate, while other disasters like the COVID-19 pandemic \citep[e.g.,][]{FGL21,WSL23} can elevate it. Therefore, the ensuing extrapolations on multi-decadal timescales must be interpreted with appropriate caution.

\begin{figure}
\includegraphics[width=7.7cm]{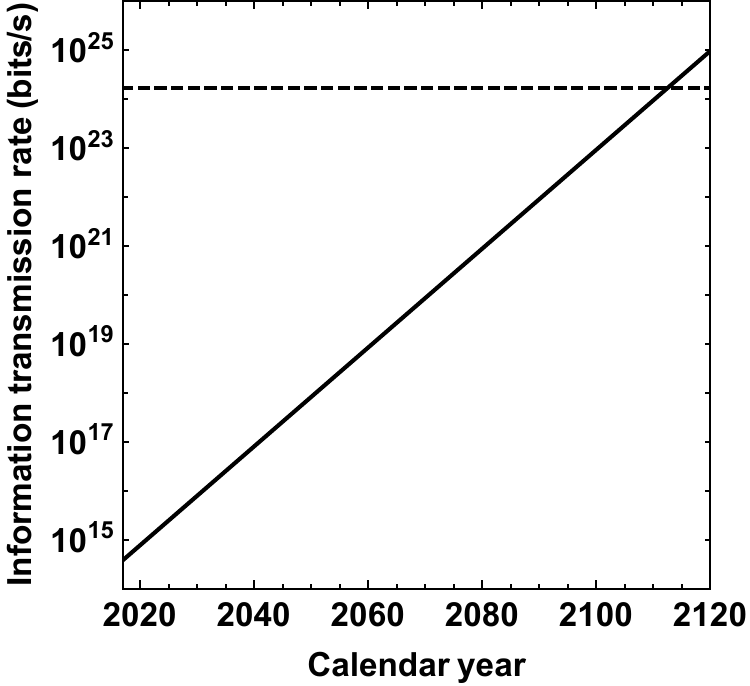} \\
\caption{Information transmission rate associated with communication (in bits/s) as a function of the calendar year. The dashed line is the rate estimated for the biosphere (assuming it is roughly constant on short timescales) and the solid line signifies the rate for the technosphere given by (\ref{Itdef}).}
\label{FigITRate}
\end{figure}

By combining the above data concerning the 2017 IP traffic with the past and expected growth rate, we have
\begin{equation}\label{Itdef}
    I_t \approx 3.8 \times 10^{14}\,\mathrm{bits/s}\,\exp\left(\frac{\mathcal{T} - 2017}{4.3\,\mathrm{yr}}\right),
\end{equation}
where $\mathcal{T}$ represents the calendar year. Note, however, that the model is clearly predicated upon the key assumption of sustained exponential growth, which might not be realistic as a consequence of technological, socioeconomic, or ecological factors, along the lines articulated in the prior paragraph.

It is worth recognizing that $I_p$ is also subject to evolution, albeit on geological timescales, owing to which we treat it as effectively constant. However, if we were to travel a few Gyr into the past \citep{HKH20} or $\sim 1$ Gyr into the future \citep{OR21}, the net primary productivity of Earth would be conceivably a few percent of the current value, intimating that $I_p$ might be proportionally diminished. A precise estimate is difficult because the total biomass is not adequately constrained in those eras; current studies suggest that it may have been at least an order of magnitude lower a few Gyr ago relative to the present day \citep{MP18}.

As per the above ansatz, the value of $I_t$ in 2022 is $\sim 9$ orders of magnitude smaller than our previous estimate for $I_p$. If (\ref{Itdef}) still remains valid in the future, then $I_t$ is forecast to exceed $I_p$ in the vicinity of the year 2113 (i.e., about $90$ years from now), as evidenced by Figure \ref{FigITRate}. It is intriguing that the information content (measured in bits) of the technosphere may surpass that stored in DNA (viz., the biosphere) approximately $100$ years in the future \citep{GHK18}, which is nearly equal to the timescale calculated herein.

\section{Discussion and Conclusions}\label{SecConc}

Our analysis, while indubitably simple, yields many crucial implications. First and foremost, the global rate of information transmission in the biosphere might be potentially $\sim 9$ orders of magnitude higher compared to the technosphere in 2022.\footnote{In contrast, the power requirements for communication in the biosphere need not be commensurately higher than the equivalent for the technosphere, given that the thermodynamic efficiency of cellular computations is $\sim 6$ orders of magnitude greater relative to electronic computing devices \citep{KWC17}.} In turn, this prediction indicates that the biosphere is significantly more ``communication-rich'', broadly speaking, with respect to the technosphere, although this status quo may undergo transformation in less than a century if the latter exhibits continuous exponential growth due to rapid technological advancements.

Second, this work foregrounds the informational aspects of life, and demonstrates that Earth's complex biosphere is distinguished by substantial information flow. Hence, traditional yardsticks such as energy \citep[e.g.,][]{VS08,VS10} or energy rate density \citep{EJC01,EJC11} utilized to categorize complex systems may be supplemented by information-centric metrics like the information transmission rate employed herein. For instance, the Kardashev scale in astrobiology \citep{NK64}, which classifies putative technological species (including humans) based on their energy consumption (i.e., power), could be complemented by a scheme centered on the information transmission rate, or alternative measures of information content \citep{WCZ16}.

Lastly, this line of inquiry opens up novel vistas for further research. To conduct an apples-to-apples comparison, we have restricted ourselves to assessing the global information transmission rates entailing communication. It is feasible to dig deeper into the properties of the underlying information networks and systematically analyze their topology, for example, to ascertain whether they are scale-free, which is ostensibly the case \citep{KSM19}. By the same token, in lieu of canonical (mutual) information, state-of-the-art quantitative concepts endowed with additional nuance such as semantic information \citep{KW18} or functional information \citep{HGC07} could be harnessed.

Interestingly, while our planet's biosphere has maintained a mostly stable state -- albeit punctuated by evolutionary radiations and mass extinctions \citep{RH12,DW16,KN17,MJB20,AHK21} -- over the span of Gyr \citep{JWK89,CK92,TML98,FB07,LDD18,JFK19}, the physical facets of the technosphere (engendered by humans) witnessed explosive growth in the Anthropocene \citep{ZWW17}, to wit, on a timescale of $\lesssim 100$ years. In a similar vein, our work suggests that $I_t$ (embodying the informational aspects of the technosphere) is roughly experiencing exponential growth currently, which might potentially lead to this quantity approaching and/or overtaking $I_p$ is less than a century from now. Understanding the information architecture of Earth's biosphere and technosphere (e.g., relative amounts of semantic and syntactic information) may enable us to go beyond the coarse-grained metric of the global information transmission rate introduced in this paper, and thereby illuminate the differences between the biosphere and technosphere (and their trajectories). This point might prove to be important, as the detrimental effects of the Anthropocene, vis-\`a-vis long term negative impacts, may conceivably fall more strongly on the technosphere compared to the biosphere.

We round off our exposition by reiterating the major caveats: some of the model parameters are poorly constrained, and this simple order-of-magnitude treatment may not account for certain vital processes. Hence, our approach should be viewed as heuristic, and as a preliminary foray into this subject that can, in principle, facilitate and/or stimulate subsequent research. Of the multiple parameters in our approach, we highlight the need to perform in-depth analyses of the fraction of time that cells communicate in biofilms and the average information transmission rate between each pair of cells in such environments (as mentioned in Section \ref{SSecITBio}). Constraining these variables will help us gain a more reliable understanding of the magnitude of $I_p$.

Nonetheless, in spite of these limitations, it seems credible that our central conclusion is robust, namely, that information transmission via communication in the biosphere is many orders of magnitude higher than the current technosphere. If this proposition is indeed correct, the biosphere apparently comprises the dominant component of communication when we seek to either comprehend the present or gaze backward in time, but the future trajectory of this phenomenon might be indelibly sculpted by the technosphere in the Anthropocene and beyond.

\section*{Author contributions}
Conceptualization, M.L., A.F. and A.B.; methodology, M.L.; formal analysis, M.L.; writing---original draft preparation, M.L.; writing---review and editing, A.F. and A.B.; supervision, A.F. and A.B.; project administration, M.L.; funding acquisition, A.F. All authors have read and agreed to the published version of the manuscript.

\section*{Funding}
The research undertaken by M.L. and A.F. was partly funded by the NASA Exobiology program under the grant 80NSSC22K1009. M.L. acknowledges fiscal support from the Florida Tech Open Access Subvention Fund.

\section*{Institutional review}
Not applicable.

\section*{Informed consent}
Not applicable.

\section*{Data availability}
No new data were created or analyzed in this study. Data sharing is not applicable to this article. 

\section*{Acknowledgements}
M.L. is grateful to Sandeep Choubey and Chris McKay for perspicacious comments regarding a cognate manuscript.

\section*{Conflicts of interest}
The authors declare no conflict of interest.

\bibliographystyle{abbrvnat}
\bibliography{GlobalIT}

\end{document}